\documentclass[a4paper]{jpconf}
\usepackage{graphicx}
\begin{document}
\title{Indirect methods in nuclear astrophysics}

\author{C.A. Bertulani$^{(a,b)}$, Shubhchintak$^{(a)}$, A. Mukhamedzhanov$^{(b)}$, A. S. Kadyrov$^{(c)}$, A. Kruppa$^{(d)}$, and D. Y. Pang$^{(e)}$}

\address{$^{(a)}$Department of Physics and Astronomy, Texas A\&M University-Commerce, Commerce, TX 75429, USA}

\address{$^{(b)}$Department of Physics and Astronomy, Texas A\&M University, College Station, TX 75429, USA}

\address{$^{(c)}$Department of Imaging \& Applied Physics, Curtin University, GPO Box U1987, Perth 6845, Australia}

\address{$^{(d)}$Institute for Nuclear Research, Hungarian Academy of Sciences, Debrecen, PO Box 51, H?4001, Hungary}

\address{$^{(e)}$School of Physics and Nuclear Energy Engineering, Beihang University, Beijing 100191, China}

\address{}

\ead{carlos.bertulani@tamuc.edu, shub.shubhchintak@tamuc.edu}

\begin{abstract}
We discuss recent developments in  indirect methods used in nuclear astrophysics to determine the capture cross sections and subsequent rates of various stellar burning processes, when it is difficult to perform the corresponding direct measurements. We discuss in brief, the basic concepts of Asymptotic Normalization Coefficients, the Trojan Horse Method, the Coulomb Dissociation Method, (d,p), and charge-exchange reactions.
\end{abstract}

\section{Introduction}
In astrophysics there is a plethora of interesting questions related to the evolution of the Universe, of processes responsible for energy generation in stars and how the nuclei that are observed on Earth are formed. Particularly, one is concerned with various processes in the pp-chain, CNO-cycle, $r$-process, $s$-process, etc., through which elements are created in the Universe. The knowledge of cross sections of some specific nuclear reactions such as capture reactions (${\rm p},\gamma$), (${\rm n}, \gamma$), ($\alpha, \gamma$) give important information about these processes.

Although direct experiment are preferable, in a large number of cases they are difficult to carry out. This is mainly due to the following reasons:
\begin{enumerate}
\item Experiments have to be performed at stellar energies which are usually very small (of the order of tens or hundreds of keV/u). The cross sections for the reactions involving charged particles are very small (nano barns or pico barns). Often, even after long hours of data collection only a few events are obtained. Background and stability problems have to be taken care of \cite{Aliotta}. Direct measurements at low energies can also be affected by electron screening, requiring a difficult and uncertain treatment. An alternative way is to perform experiments at higher energies (few MeV/u) and then extrapolate the results down to the desired energies. But this procedure also involves considerable uncertainties.
\item Many of the stellar reactions involve unstable nuclei with very short life time.
\end{enumerate}

\section{Indirect methods}
To mitigate these problems, alternate indirect methods (a combination of experimental and theoretical analysis) are currently  being used where equivalent information can be extracted by performing reactions at much higher beam energies. The indirect methods depend upon the type of reaction involved. Of these, three indirect methods stand out  in present day nuclear astrophysics: (a) Asymptotic Normalization Coefficient method,  (b) Trojan horse method and  (c) Coulomb dissociation method. See, e.g., the review articles \cite{akram_carlos,Gade}.

\subsection{\bf Asymptotic Normalization Coefficient}
The Asymptotic Normalization Coefficient (ANC) method is based on the normalization of the tail of the quantum overlap of bound state wave functions of the initial and final nuclei \cite{akram_carlos,hmxu}. Capture reactions in stellar environments take place either by direct capture or by resonant capture. Direct capture reactions of charged particles usually involve systems with small binding energies and they are mostly peripheral at stellar energies. On the other hand, neutron capture reactions may contain contributions from the nuclear interior and are very sensitive to the spectroscopic factor (SF) of the final state \cite{ak2008}. The presence of the Coulomb barrier for charged particle capture causes the reaction to be peripheral. Classically, if the incident particle energy is below the barrier then no capture will occur, however quantum mechanical `tunneling' gives a probability of barrier penetration. Therefore, the capture in such case proceeds through the tail of the nuclear overlap of the initial and final bound state wave functions whose shape is completely determined by the Coulomb force and the amplitude of the tail is given by the ANC. Although the ANC method has been mostly used for charged particle direct capture reactions,  it can also be used for neutron capture (${\rm n},\gamma$) reactions, as explained in Ref. \cite{ak2008}. 
One can extract the ANC from differential cross section data of peripheral particle transfer reactions having the same vertex or from the one-nucleon breakup reaction of loosely bound nuclei. 

Consider the virtual decay process $B \rightarrow A + a$. If $\phi_B(\zeta_a,\zeta_A;{\bf r})$, $\phi_a(\zeta_a)$ and $\phi_A(\zeta_A)$ are the bound state wave functions for respective nuclei with $\zeta$'s being the internal coordinates, then the overlap function is given by
$
I^{B}_{Aa}({\bf r})=\left<\phi_A(\zeta_A)\phi_a(\zeta_a)|\phi_B(\zeta_a,\zeta_A;{\bf r})\right>. \label{a1}
$
In the asymptotic limit (when the nuclear forces are vanishingly small), the radial part $I^{B}_{Aa}({r})$ of the overlap function can be expressed in terms of Whittaker function ($W_{a,b}$),
\begin{eqnarray}
I_{Aa}^{B}(r)= C_{Aa\ell_Bj_B}^{B}\frac{W_{-\eta,\ell_B+1/2}(2kr)}{r};     \hspace{0.4in} r > R_N,   \label{a2}
\end{eqnarray}
where $R_N$ is the nuclear interaction radius of $A$ and $a$, $C$ is the ANC. It defines the amplitude of the tail of the radial overlap integral and it depends upon the structure of nucleus $B$. $k$ and $\eta$ are the wave number and the Coulomb parameter for the bound state $(Aa)$, respectively. $\ell_B$ and $j_{B}$ are the orbital and total angular momentum of $B$.

The above formalism of ANC is based on the overlap integral and is also called Schr\"{o}dinger formalism as it involves wave functions and potentials.
However, there is another way to incorporate ANC in the theory of direct reactions. This formalism is based on general scattering theory and it is valid both in non-relativistic quantum mechanics as well as in field theory. Here, the ANC can be used to find the residues at the poles (for bound or resonance states) of the $S$-matrix for elastic scattering of the $a + A$ system. For more details about this method, see Ref. \cite{akram_carlos}.
The  cross section for the direct radiative capture reaction $A + a \rightarrow B + \gamma$ can be written as
\begin{eqnarray}
\sigma(E_a) = KF\left |\left< I_{Aa}^{B}(r)\left|\hat{O}(EL)\right|\chi_i^{(+)}(r)\right>\right|^2 , \label{a3}
\end{eqnarray}
where $KF$ is a kinematic factor, $\chi_i^{(+)}$ is the scattering wave function in the initial channel and $\hat{O}(EL)$ is the electromagnetic transition operator connecting the initial and final channel and it does not involve the structure of the core $A$. 
In the asymptotic limits the capture cross section [Eq. (\ref{a3})] can be written as
\begin{eqnarray}
\sigma(E_a) = KF \left|\left<C_{A a \ell_B j_B}^{B}{W_{-\eta,\ell_B+1/2}(2kr)}/{r}\left|\hat{O}(EL)\right|\chi^{(+)}(r)\right>\right|^2 
= (C_{A a \ell_B j_B}^{B})^2 w(E_a),  \label{a4}
\end{eqnarray}
where $w(E_a)$ is a function independent over the bound state structure of $B$. Thus,  the direct capture cross section completely depends upon the ANC. By determining the ANC from some other direct reaction experiment (e.g., from elastic scattering), one can calculate the capture cross section at stellar energies. However, one should keep in  mind the necessary conditions  that the reaction chosen for extracting the ANC should have the same vertex and also that it should be peripheral in nature. The common choices are (i) a transfer reaction between the heavy ions, having one of the vertices kept the same as the vertex for the desired direct capture reaction and (ii) a breakup reaction of loosely bound nuclei. 

\begin{figure*}[ht]
\begin{center}
\includegraphics[trim=2.5cm 3.5cm 2.5cm 1.8cm,clip,width=7.5cm]{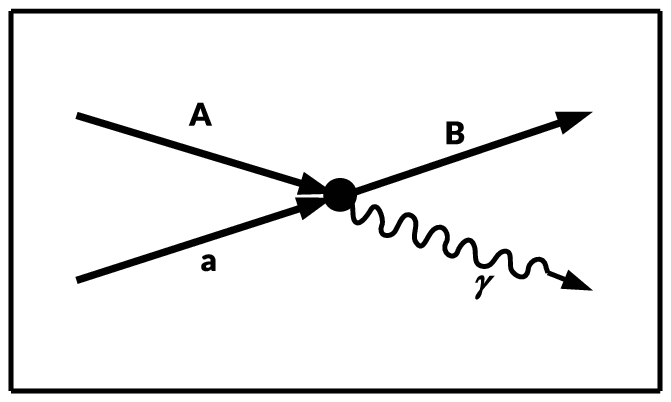}
\includegraphics[trim=2.5cm 3.5cm 2.5cm 1.8cm,clip,width=7.5cm]{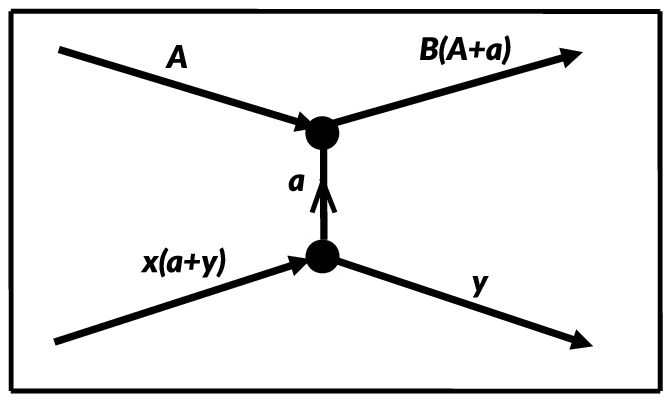}
\caption{(a) The reaction vertex in the radiative capture reaction $A(a, \gamma)B$ and (b) the reaction vertices in the transfer reaction $A(x, y)B$.}
\label{fig1}       
\end{center}
\end{figure*}

\noindent{\bf{\underline{ANC from transfer reactions}}}.
Consider the transfer reaction $A(x, y)B$, with $x = y + a$ and $B = A + a$. Figure \ref{fig1} shows the vertex (vertices) for the direct radiative capture (transfer) reaction. The distorted-wave Born approximation (DWBA) is used to analyze such type of peripheral transfer reactions. The DWBA cross section ($\sigma^{DW}$) for the transfer reaction can be related with the measured angular distribution of the nuclei by means of the relation 
\begin{eqnarray}
\frac{d\sigma}{d\Omega} = (S_{A a \ell_B j_B}^{B}) (S_{y a \ell_x j_x}^{x}){\sigma_{\ell_B j_B \ell_x j_x}^{DW}}, \label{a5}
\end{eqnarray}
where $S^{i}$, $\ell_{i}$ and $j_{i}$ ($i = B$ and $x$) are the SF, and the orbital and total angular momentum of the respective nuclei. These SF's are related to the corresponding ANC's of the overlap function by
\begin{eqnarray}
(C_{A a \ell_B j_B}^{B})^2 = (S_{A a \ell_B j_B}^{B})^2  (b_{A a \ell_B j_B}^{B})^2    ;       \hspace{0.4in} (C_{y a \ell_x j_x}^{x})^2 = (S_{y a \ell_x j_x}^{x})^2 (b_{y a \ell_x j_x}^{x})^2,  \nonumber
\end{eqnarray}
where $b$'s are the single particle ANC's of the particle $a$ in the nuclei $x$ and $B$ whereas, $C^x$ and $C^B$ are the ANC's of the initial and final nuclei. Therefore, Eq. (\ref{a5})  can be written in terms of ANC's as
\begin{eqnarray}
\frac{d\sigma}{d\Omega} = (C_{A a \ell_B j_B}^{B})^2 (C_{y a \ell_x j_x}^{x})^2\frac{\sigma_{\ell_B j_B \ell_x j_x}^{DW}}{b_{A a j_B \ell_B}^2b_{y a \ell_x j_x}^2}. \label{a6}
\end{eqnarray}

When more than one orbital participate, a summation over all the orbitals is required. For  peripheral reactions, the ratio in Eq. (\ref{a6}) is independent of the single particle ANC's ($b$'s in the denominator), which also gives a criterion for peripheral reactions.   
Therefore, by comparing the DWBA cross section with the experimental result one can extract the spectroscopic factor and hence the ANC ($C_{A a \ell_B j_B}^{B}$) for the vertex of interest ($A + a \rightarrow B$), provided that the information about the other vertex ($x \rightarrow y + a$) (i.e. $C_{y a \ell_x j_x}^{x}$) is already known from other independent measurements. This method has been tested in several direct capture reactions to obtain astrophysical S-factors \cite{Muk15}.

\noindent{\bf{\underline{ANC from the single nucleon removal reactions}}}. It is a well known fact that  loosely bound nuclei have one or two nucleon(s) well detached from core nucleons and that they can be easily removed in the nuclear field of a target. Such reactions are usually peripheral even at energies of few hundred MeV/nucleon and they are dominated by the asymptotic tail of the removed nucleon wave function. From the measurement of momentum distributions of the core, one can extract the spectroscopic information about the relative orbital angular momentum of the fragments and contribution of the different single particle states \cite{gregers}. From the measured one-nucleon removal cross section one can also extract the ANC for the capture process of interest \cite{trache}.
The one-nucleon removal cross section ($\sigma$) for the reaction $B \rightarrow A + a$ can be written as
$
\sigma =  {(C^B_{A a \ell j})^2 \sigma_{sp}}/{(b^B_{A a \ell j})^2}, \label{a7}
$
where $\sigma_{sp}$ is the calculated single particle removal cross section. One has to take the sum over all the orbitals contributing to the total cross section. The advantage of this technique is that the one-nucleon removal cross sections are usually large even at high energies and therefore a beam of lesser quality can be used.
 
\subsection{\bf Trojan Horse method}
This is an another indirect method, where the cross section of a suitable two-body to three-body reaction ($A + x \rightarrow c + B + y$) is used to extract the cross section for a relevant two-body reaction ($A + a \rightarrow c + B$) at stellar energies. This specific type of reaction (two-body to three-body) is called Trojan Horse (TH) reaction and hence the method is known as the Trojan Horse Method (THM) \cite{baur}. A TH reaction is chosen in such a way that nucleus $x$ contains $a$ in its cluster structure ($x = y + a$) and this cluster $a$ interacts with $A$, whereas the other cluster $y$ remains as a spectator. For this to happen, one has to keep Quasi-Free (QF) kinematical conditions for the clusters $y$ and $a$, which is based on the consideration that the relative momentum between these clusters in the three-body phase space should be zero or small compared to the wave number of bound state ($ay$) \cite{spitaleri}. In that case the interaction between $y$ and $a$ would be small and hence they will stay at maximal distance from each other. In the above reaction if $c$ stands for $\gamma$ then it represents a radiative capture reaction. In fact, this method is very flexible and can be used to study many other type of reactions and is not only limited to the radiative processes, such as the Coulomb dissociation and  the ANC method. While performing an experiment with a TH reaction, one has to pay attention to all the other possible reactions which can be triggered and that can act as background with respect to the desired reaction event.

\begin{figure*}[ht]
\begin{center}
\includegraphics[trim=2.5cm 3.5cm 2.5cm 1.8cm,clip,width=8cm]{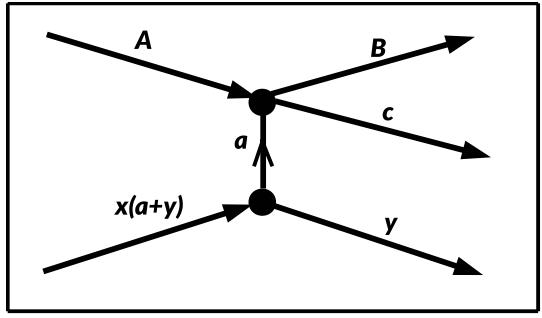}
\caption{Trojan horse reaction, with three particles in the final state}
\label{fig2}       
\end{center}
\end{figure*}

As mentioned earlier, the direct measurement of a capture cross section for the binary process $A + a \rightarrow c + B$ at stellar energies is difficult because one has to take into consideration the electron screening effect  and  the extrapolation to low energies is not reliable because of unexpected resonances \cite{Gade}. However, one can use the THM to obtain cross sections down to the relevant ultra-low stellar energies. This is because the TH reaction between $x$ and $A$ takes place at energies above the Coulomb barrier so that $A$ can dig into $x$ and there is no possibility of additional an Coulomb barrier between $A$ and $a$. Here, $a$ acts as a virtual particle and hence does not obey the energy-momentum relations for a free particle. Therefore, the cross sections of the binary sub-process extracted from the TH processes do not contain the related Coulomb barrier and are free from electron screening. Consequently, with this method one can determine the astrophysical S-factor S(E) of the binary process $A(a,c)B$ at low relative kinetic energies of the particles $a$ and $A$ without being affected by electron screening. In addition, comparing the direct capture cross section with those obtained from the THM, one can determine the screening potential.

As explained in Ref. \cite{tumino},  in spite of the high energy ($E_A$) of a projectile $A$ (above the Coulomb barrier of $A+x$), the binary reaction $A+a$ still takes place at low relative energies $E_{Aa}$ because of the QF condition. Figure \ref{fig2} describes the mechanism for the THM. If $\epsilon_{ay}$ is the binding energy of intercluster $a-y$ in nucleus $x$ then the relative energy of $A-a$ is given by
\begin{eqnarray}
E_{Aa}=\frac{p^2_{Aa}}{2\mu_{Aa}}-\frac{p^2_{ay}}{2\mu_{ay}}-\epsilon_{ay}, \label{a8}
\end{eqnarray}
where $p$'s and $\mu$'s are the relative momenta and reduced masses of the respective pair. With the QF kinematics, $p_{ay} = 0$, and also considering target $x$ at rest (i.e. momentum $P_x = 0$) in the lab system, the relative energy $E_{Aa}$ from Eq (\ref{a8}) is
\begin{eqnarray}
E_{Aa}=\frac{m_a}{m_a+m_A}E_A-\epsilon_{ay}, \label{a9}
\end{eqnarray}
with $m$'s being the masses of respective nuclei. Thus, it is clear that due to the presence of the factor ${m_a}/{(m_a+m_A)}$ and the binding energy $-\epsilon_{ay}$, the relative energy $E_{Aa}$ for the binary reaction remains very small and even negative in spite of high $E_A$.

As mentioned earlier, in the QF kinematics, there is negligible interaction of the spectator $y$ with $B$ and $c$ and hence in the final three-body state only the scattering of two fragments $c$ and $B$ is important. Therefore, the plane-wave-impulse approximation (PWIA) is used to analyze the experimental results in the THM. In this approximation, the three-body triple differential cross section for the TH reaction is factorized as \cite{tumino,pizzone}
\begin{eqnarray}
\frac{d^3\sigma}{dE_{c}d\Omega_{c}d\Omega_{B}}= KF|\phi_x(k_{ay})|^2 \Big(\frac{d\sigma^{HOES}}{d\Omega_{c.m.}}\Big), \label{a10}
\end{eqnarray}
where
\begin{enumerate}
\item $KF$ is the kinematical factor which contains the final state phase space factor in terms of masses, angles and momenta of outgoing particles (see Ref. \cite{tumino}). 
\item $\phi_x(k_{ay})$ is the Fourier transform of the bound state wave function of $x$ which describes the momentum distribution of $x$ at the final momentum of $a-y$ intercluster. For a nucleus $x$ having ground state angular momentum $\ell = 0$, the application of this method is much simpler as the Fourier transform of the ground state wave function peaks at zero momentum and is well known. This helps in determining the QF scattering angles. On the other hand, for high-$\ell$ nuclei, the momentum distribution becomes much broader and hence the internal separation of the fragments $a-y$ will be small and $y$ can no longer be treated as spectator. Therefore, high-$\ell$ ($>0$) nuclei are less suitable for THM (see Ref. \cite{spitaleri}). 
\item ${d\sigma^{HOES}}/{d\Omega}$ is the half-off-energy-shell (HOES) differential cross section for the two body reaction $A + a \rightarrow B + c$ at relative kinetic energy $E_{QF}$ of the $A-a$ system, such that $E_{QF} = E_{cB}-Q$. Here, $E_{cB}$ is the kinetic energy of the relative motion of $c$ and $B$ and $Q$ is the $Q$-value of the binary reaction. 
\end{enumerate}
It is clear from Eq. (\ref{a10}) that after calculating the factors $KF$ and $|\phi_x(k_{ay})|^2$, the HOES binary cross section can be obtained from the measured triple differential cross section of the TH reaction as
\begin{eqnarray}
\frac{d\sigma^{HOES}}{d\Omega_{c.m.}}\propto\frac{d^3\sigma}{dE_{c}d\Omega_{c}d\Omega_{B}}\frac{1}{KF|\phi_x(k_{ay})|^2 } . \label{a12}
\end{eqnarray}
This binary HOES cross section is then used to extract the quantity of interest, namely, the on-energy-shell cross section by normalization to the direct data available at higher energies. The THM has been used in many theoretical and experimental studies of capture reactions with success. Although, initially it has been used for charged particle capture reactions, now it has also been extended to study direct neutron and resonant capture reactions. 

\subsection{\bf Coulomb dissociation method} 
\begin{figure*}[ht]
\begin{center}
\includegraphics[trim=2.5cm 3.5cm 2.5cm 1.8cm,clip,width=8cm]{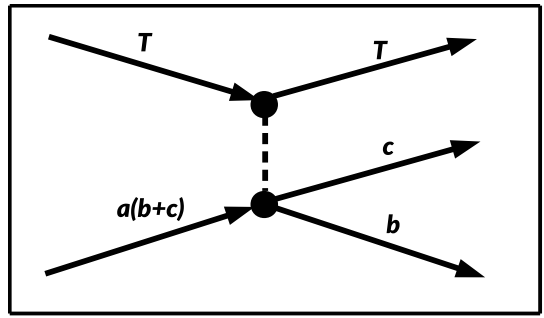}
\caption{Coulomb breakup of nucleus $a$ in the electromagnetic field of heavy target $T$.}
\label{fig3}       
\end{center}
\end{figure*}

The Coulomb dissociation/breakup method (CD) is an another indirect method proposed by Baur {\it et al} \cite{baur_carlos}. As Coulomb induced dissociation is an inverse to the radiative capture process, one can relate the measured Coulomb dissociation cross section (which can be measured at high beam energies and are larger in magnitude) to the relevant radiative capture cross section of astrophysical interest. 
In this method the composite projectile $a$ consisting of two sub-structures $b$ and $c$, is  broken up in the Coulomb field (or electromagnetic field) of a heavy target nucleus $T$ (see Fig. \ref{fig3}), 
$
a + T \rightarrow b + c + T, \label{a12}
$
where $b$ is the core nucleus and $c$ is a valence nucleon which can be charged or uncharged. The advantage of the CD method is that the measurements are  performed at even higher beam energies and the cross sections are considerably larger as the energy increases.  Also at sufficiently high incident velocities the fragments $b$ and $c$ emerge with high energies which in turn facilitates their detection. With adequate kinematical conditions in coincidence measurements the study of low relative energies of the final state fragments is still possible \cite{baur_carlos}. 
However, for the viability of the method one has to be ensure that the nuclear breakup effect should be negligible or should be known accurately. For a pure Coulomb breakup, a large impact parameter or small scattering angle is often required. Furthermore, the breakup process is considered to be elastic, so that there will be no excitations in the target, although this condition can be relaxed if the energy transferred to the excitation is small compared to the bombarding energy, what is usually the experimental choice. 

The Coulomb breakup process [Eq. (\ref{a12})], is then related to the photodisintegration reaction $a + \gamma \rightarrow b + c$, where one calculate the photodisintegration cross sections ($\sigma_{\gamma, n}^{\pi\lambda}$) as
\begin{eqnarray}
\frac{d\sigma}{dE_{rel}}=\frac{1}{E_{\gamma}}\sum_{\pi\lambda}\sigma_{\gamma, n}^{\pi\lambda}\hspace{0.1cm}{\rm n}_{\pi\lambda},\label{a13}
\end{eqnarray}
where ${d\sigma}/{dE_{rel}}$ is the relative energy spectra in the Coulomb breakup at relative energy $E_{rel}$ of $b-c$ system, $\pi$ stands for either electric or magnetic transition of multipolarity $\lambda$, $E_{\gamma}=E_{rel}+Q$ is the photon energy with ${\it Q}$ as the ${\it Q}$-value of the reaction and ${\rm n}_{\pi\lambda}$ is the equivalent photon number (also called virtual photon number) and depends on the $a - t$ system (see Ref. \cite{Bertulani}).  Similar to Eq. (\ref{a13}) the angular differential cross section of the Coulomb breakup can also be related to photo disintegration cross section as
\begin{eqnarray}
\frac{d\sigma}{d\Omega}=\frac{1}{E_{\gamma}}\sum_{\pi\lambda}\sigma_{\gamma, n}^{\pi\lambda}\hspace{0.1cm}\frac{{d\rm n}_{\pi\lambda}}{d\Omega},\label{a14}
\end{eqnarray}
$\Omega$ being the scattering solid angle.

It has been found that relativistic effects, higher order and higher multipole's contribution can influence the reaction process \cite{bertulani_rel,bertulani_multi} and hence make it difficult to use Eqs. (\ref{a13})-(\ref{a14}). One has to take these effects into account while using the CD method. When a single multipolarity of either type (electric or magnetic) dominates, the above equations can be easily used to find the photodisintegration cross section for the dominating multipolarity ($\lambda$). Using the `principle of detailed balance' the capture cross section ($\sigma_{n, \gamma}$) for the desired capture process $b + c \rightarrow a + \gamma$, can be obtained from $\sigma_{\gamma, n}^{\pi\lambda}$ as
\begin{eqnarray}
\sigma_{n, \gamma} = \frac{2(2j_a+1)}{(2j_b+1)(2j_c+1)}\frac{k_{\gamma}^2}{k^2_{bc}}\sigma_{\gamma, n}^{\pi\lambda}, \label{a15}
\end{eqnarray}
where $j_a$, $j_b$ and $j_c$ are the spins of  particles $a$, $b$ and $c$, respectively. $k_\gamma$ and $k_{bc}$ are the wave numbers of the photon and that of relative motion between $b$ and $c$, respectively. As a word of caution, the principle of detailed balance cannot be used for the situations where capture leads to the bound excited states having unknown branching ratios of $\gamma$-ray emission \cite{moto}.

The CD method has been used to determine the capture cross sections for numerous radiative capture stellar processes using both experimental and advanced theoretical studies (see Ref. \cite{akram_carlos}). In fact, there are different theoretical models used to study the Coulomb dissociation. In Refs. \cite{pb,shub1,shub2}, a pure quantum mechanical model \cite{rc} using the post-form DWBA is used to calculate the radiative neutron capture cross section on $^{8}$Li, $^{14}$C and $^{15}$N from the Coulomb breakup of $^{9}$Li, $^{15}$C and $^{16}$N, respectively. For $^{15}$N($n,\gamma$)$^{16}$N, there are large contributions from the capture to the low-lying bound excited states of $^{16}$N. Therefore, although theoretically such a study is possible, experimentally it is difficult as it requires additional information on the $\gamma$-ray branching ratios for the excited states, similar to what was done in Ref. \cite{izsak}. Recently, in Refs. \cite{shub3,shub4}, the theory of Coulomb dissociation \cite{rc} (for neutron exotic nuclei) has been extended to include the projectile deformation. This further opens the door to study the affect of deformation on the neutron capture cross sections and subsequent rates for medium mass nuclei. 

\section{\bf (d,p) and charge-exchange reactions}
 In surrogate reactions such as  (d,p) reactions,  the neutron is brought to react with a nucleus inside a Trojan, or surrogate nucleus (here, the deuteron).  Such reactions are thought to give information on neutron-induced reactions, often not available in direct measurements.  The neutron inside the deuteron has angular momentum and other quantum numbers that are not equal to those of free neutrons at the same energy. If angular momentum is not relevant in the reaction of interest (a rare situation), the desired neutron-induced reaction can be extracted from the surrogate equivalent. (d,p) reactions  are not as simple as one might initially think because they involve at least three particles in the incoming and outgoing channels. Assuming it can be treated as a three-body reaction, one may use the {\it Alt-Grassberger-Sandhas} (AGS) \cite{AGS67} formalism to describe the reaction.  The AGS equations are a branch of the Faddeev formalism to treat the $(a+b) + A$ system, and its rearrangements $a + (b+A)$ and $b+ (a+A)$. The three-particle scattering in each channel is obtained from the transition operators $T_{\beta\alpha}$, where $\alpha$($\beta$) corresponds to channel permutation combinations. They are the set of coupled integral equations 
\begin{equation}
T_{\beta\alpha}(z)=\tilde{\delta}_{\alpha\beta} G_0^{-1}(z)+\sum_{\nu=1}^3\tilde{\delta}_{\beta\nu}t_\nu(z) G_0(z)T_{\nu\alpha}(z) \ ,
\end{equation}
where  $\tilde{\delta}_{\beta\nu} =1-\delta_{\beta\nu}$ is the anti-delta-Kronecker symbol, and $G_0 =(E+i_{\epsilon^+}-H_0)^{-1}$ for the three-free particle c.m. energy $E$ and Hamiltonian $H_0$. The two-body transition operator $t_\nu$ for each interacting pair with inter-potential $v_\nu$ is determined from the Lippmann-Schwinger equation
$t_\nu=v_\nu+v_\nu G_0 t_\nu.$
Additional approximations are often necessary and the correct treatment of the Coulomb interaction poses a major difficulty to solve the AGS set of equations \cite{Muk14}. One can treat the Coulomb interaction with help of screened Coulomb potentials, but this method works only for light nuclei since wild oscillations in the asymptotic region jeopardizes the accuracy of the solutions \cite{DF09}. 

In charge-exchange reactions, such as (p,n) reactions, one seeks information on  Gamow-Teller matrix elements which cannot be studied in $\beta$-decay experiments. The method is useful to obtain cross sections for inelastic neutrino scattering and electron capture reaction rates in stars. Not only  (p,n), but ($^3$He,t) and other charge exchange reactions at energies around 100 MeV/nucleon have been used \cite{Sas12}. 
Experimentally, one often uses simplified theories in the experimental analysis, e.g., the cross section $\sigma(p,n)$ at small momentum transfer $q$ is assumed to be \cite{Taddeucci1987}, 
\begin{equation}
{d\sigma\over dq}(q=0)=KF.N_D|J_{\sigma\tau}|^2 B(\alpha) , \label{tadeucci}
\end{equation}
where $KF$ ($N$) is a kinematical (distortion) factor,  approximately describing  initial and final state interactions, $J_{\sigma\tau}$ is the Fourier transform of the charge-exchange interaction, and $B(\alpha=F,GT)$ is the reduced transition probability for non-spin-flip, $B(F)= (2J_i+1)^{-1}|
\langle f ||\sum_k  \tau_k^{(\pm)} || i \rangle |^2$, and spin-flip, $B(GT)= (2J_i+1)^{-1}| \langle f ||\sum_k \sigma_k \tau_k^{(\pm)} ||
i \rangle |^2$, transitions.  The simplified approach based on Eq. (\ref{tadeucci}) has been criticized in other theoretical treatments \cite{Ber93} and is currently under further theoretical study \cite{pang}.  

\section*{Acknowledgments}
{\small C.A.B. and A.M.M. acknowledge funding support by the NSF, PHY-1415656, and by DOE, DE-FG02-08ER41533,  DE-FG02-93ER40773 and DE-SC0004958. A.S.K. acknowledges the support by the Australian Research Council. A.K. acknowledges funding support from Hungarian Scientific Research Fund OTKA K112962 and D.Y.P.  acknowledges support of the National Natural Science Foundation of China (Grants 11275018 and 11035001) and the Chinese Scholarship Council (Grant 201303070253).}

\end{document}